# Photonic Wire Bonds for Terabit/s Chip-to-Chip Interconnects

N. Lindenmann[1], G. Balthasar[1], D. Hillerkuss[1], R. Schmogrow[1], M. Jordan[1],
J. Leuthold[1,2], W. Freude[1,2], C. Koos[1,2]

[1]*Institute of Photonics and Quantum Electronics (IPQ), Karlsruhe Institute of Technology (KIT), Engesserstrasse 5, 76131 Karlsruhe Germany*

[2]*Institute of Microstructure Technology (IMT), Karlsruhe Institute of Technology (KIT), Hermann-von-Helmholtz-Platz 1, 76344 Eggenstein-Leopoldshafen, Germany*

christian.koos@kit.edu

**Photonic integration has witnessed tremendous progress over the last years, and chip-scale transceiver systems with Terabit/s data rates have come into reach [1], [2]. However, as on-chip integration density increases, efficient off-chip interfaces are becoming more and more crucial. A technological breakthrough is considered indispensable to cope with the challenges arising from large-scale photonic integration, and this particularly applies to short-distance optical interconnects [3]. In this letter we introduce the concept of photonic wire bonding, where transparent waveguide wire bonds are used to bridge the gap between nanophotonic circuits located on different chips. We demonstrate for the first time the fabrication of three-dimensional freeform photonic wire bonds (PWB), and we confirm their viability in a multi-Terabit/s data transmission experiment. First-generation prototypes allow for efficient broadband coupling with overall losses of only 1.6 dB. Photonic wire bonding will enable flexible optical multi-chip assemblies, thereby challenging the current paradigm of highly-complex monolithic integration.**

Reliable and scalable interconnect technologies are of paramount importance for industrial deployment of photonic integrated circuits. Unlike electronics, where highly sophisticated metal wire bonding is the primary method of connecting integrated circuits to the outside world, photonics cannot rely on an interconnect technology of comparable versatility. Monolithic integration is often presented as a method to avoid costly chip-to-chip interfaces in photonic systems, but the associated technological complexity is still prohibitive for many applications. An industrially viable technology for chip-to-chip interconnects could enable novel system concepts that efficiently combine the strengths of different integration platforms, e.g., by complementing large-scale photonic-electronic integration on silicon-on-insulator (SOI) substrates with active optical elements on InP.

Previous research in the field of optical input-output technologies has primarily focused on refining fiber-chip coupling. Out-of-plane coupling has been demonstrated using diffraction gratings etched into the top surface of silicon-on-insulator (SOI) waveguides [4]. Coupling performance can be improved by using bottom mirrors [4], [5], silicon overlays [6] or numerically optimized apodized grating designs [7]. In addition, such structures can act as integrated polarization splitters [8]. While grating couplers enable optical access anywhere on the chip surface, they suffer from inherent bandwidth limitations. Grating couplers have insertion losses of 1–2 dB, and 1 dB bandwidths can be between 40 nm and 50 nm [5] – [11]. The bandwidth limitations can be overcome by in-plane coupling schemes that utilize inverse tapers [12] in combination with polymer, $SiO_2$ or SiON waveguides [13] – [16]. Coupling losses of less than 1 dB can be maintained over wavelength ranges of more than 100 nm [13] – [15]. Nevertheless, the viability of in-plane fiber-chip coupling is limited by the fact that the facets of the integrated waveguides need to be close to the chip edge, whereby the interconnect density is usually dictated by the pitch of the connecting fibers. A spatial data transmission density of 8 Tbit/s/mm has recently been demonstrated by using specially designed spot

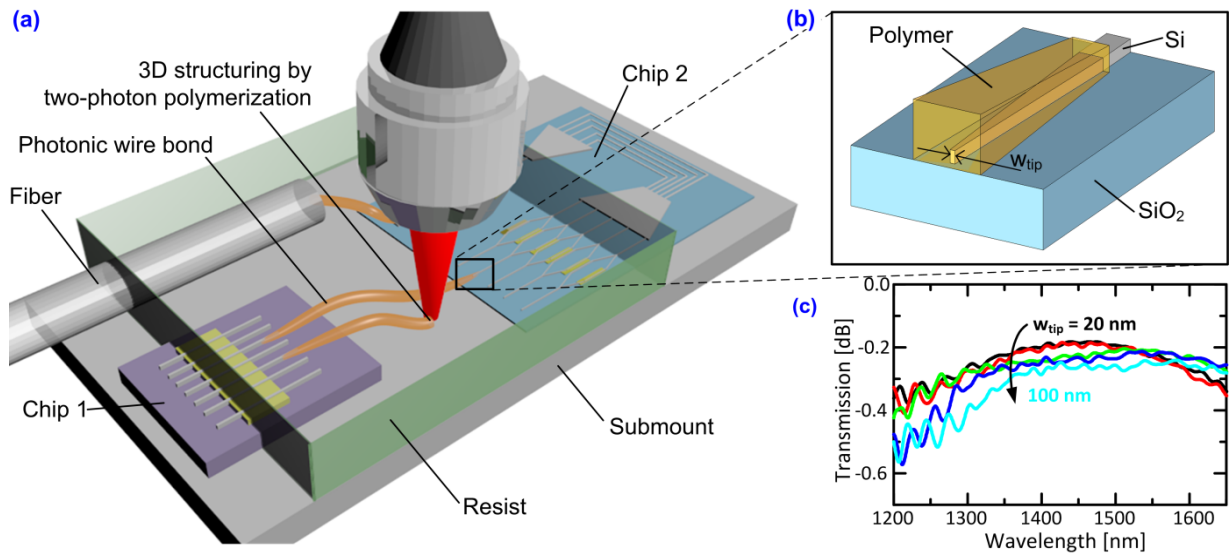

**Figure 1 (a)** Photonic wire bonding concept: Photonic chips (Chip 1, Chip 2) and optical fibers are mounted on a common submount and embedded into a photosensitive resist. Two-photon polymerization in the focus of a high-NA pulsed laser beam is then used to define three-dimensional freeform PWB structures in the volume of the resist. **(b)** Inverse-taper transition between a silicon-on-insulator (SOI) nanowire waveguide and a polymer PWB interconnect. The SOI waveguide is 500 nm wide and 220 nm high, and it is tapered down over a length of 32 μm to a tip width $w_{tip}$ between 20 nm and 100 nm, see Methods section for more details. **(c)** Simulated transition loss spectrum for different tip widths.

size converters, which interface an array of single-mode fibers with a pitch of 250 μm to an array of SOI waveguides with a 20 μm pitch [16].

However, despite these improvements, state-of-the-art interconnect techniques still rely on highly precise positioning of external fibers with respect to on-chip coupling structures. Fabrication often involves active alignment steps, where the coupling efficiency is dynamically monitored and optimized, while spatial interconnect density is usually limited by the diameter of optical fibers. Existing techniques can hardly cope with the inter-chip connectivity challenges in future photonic-electronic systems and are not well suited for efficient industrial production.

In this letter, we introduce photonic wire bonding as a novel technology for single-mode chip-to-chip interconnects [18], [17]. The technique is based on in-situ fabrication of three-dimensional (3D) freeform waveguides between prepositioned chips. The structures are created by two-photon polymerization (TPP) of negative-tone resist materials in the focus of a high-NA laser beam [19]. TPP lithography enables features sizes below the diffraction limit of the exposure wavelength [20] and has previously been used to realize functional photonic structures such as 3D photonic crystals for telecommunication wavelengths [21] or multimode optical waveguides on large-area printed circuit boards [22].

Figure 1 (a) illustrates a multi-chip system with photonic wire bonds acting as chip-to-chip and chip-to-fiber interconnects. In a first step of the fabrication process, fibers and photonic chips are fixed to a common submount using standard pick-and-place machinery with modest precision. The interconnect regions are then embedded into a photosensitive resist material, and the spatial positions of waveguide facets and coupling structures within the resist material are detected. This can be accomplished in an automated process using high-precision 3D machine vision. The shape of the PWB waveguide is adapted to these positions, so that high-precision alignment of optical devices becomes obsolete. After direct-write TPP lithography, unexposed resist material is removed, and the structures are embedded in a low-index cladding material.

Photonic wire bonding allows for both out-of-plane and in-plane coupling [17], [18], and offers a flexible way of connecting photonic chips with different waveguide technologies in an automated industrial production process. For chip-to-chip interconnects, lateral dimensions of PWB waveguides typically amount to 1 – 2 μm, and the interconnect pitch can be smaller than 5 μm. It is hence possible to accommodate hundreds of interconnects per millimeter of chip edge, or tens of thousands PWB links per square millimeter of chip surface.

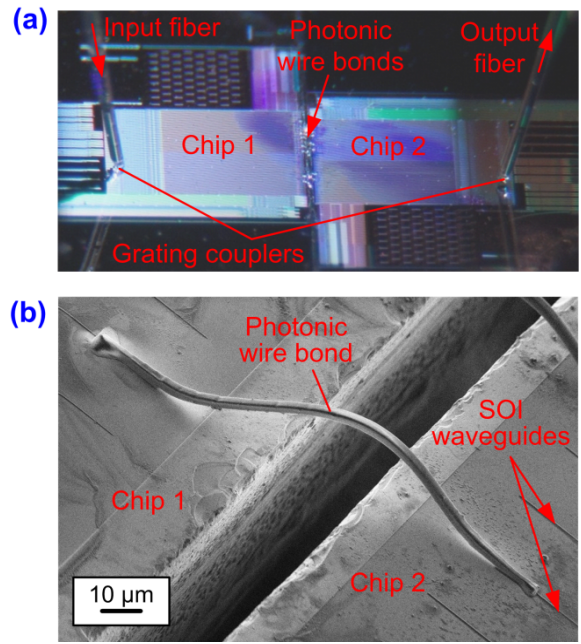

**Figure 2:** Fabricated photonic wire bonds **(a)** Optical characterization of a PWB assembly linking two SOI chips (Chip1, Chip 2). Light is coupled to the SOI waveguides using standard single-mode fibers ('Input fiber', 'Output fiber') and conventional grating couplers **(b)** PWB chip-to-chip interconnect. The structure consists of SU-8 and features a nearly rectangular cross section of approximately 2 μm by 1.6 μm. Index matching liquid was used as a low-index cladding material, residues of which are still visible on the surface of the SOI chips.

To maximize the data transmission capacity, broadband coupling between integrated waveguides and PWB interconnects is essential. For nanophotonic silicon-on-insulator (SOI) waveguides, this is achieved by inverse tapers, which are combined with three-dimensional taper structures of the PWB sections, Figure 1 (b). The coupling efficiency of the transition depends on the tip width $w_{tip}$ of the silicon taper, but even for relatively big widths of 100 nm, a broadband low-loss transmission is predicted by simulations, see Figure 1 (c). Absorption losses in the polymer waveguide of the PWB can usually be neglected: Loss figures for prevalent TPP-structurable photoresists range from 0.5 dB/cm to 3 dB/cm, whereas PWB lengths are typically less than 0.1 cm.

In a proof-of-principle experiment, we fabricated PWB interconnects between two SOI chips, Figure 2 (a). For optical characterization, light is coupled to the SOI waveguides using standard single-mode fibers and conventional grating couplers. Figure 2 (b) depicts an SEM image of the chip-to-chip interconnect waveguide. The PWB compensates for a displacement of the two SOI waveguides both in the lateral and in the vertical direction. The PWB waveguide core features a nearly rectangular cross section of approximately 2 μm by 1.6 μm and a refractive index of $n_{core}$ = 1.57 at 1550 nm, while the cladding has a refractive index of $n_{cl}$ = 1.30. Inverse SOI tapers of 30 μm length and 100 nm tip width couple the SOI waveguides to the PWB. The fabrication process is described in more detail in the Methods section.

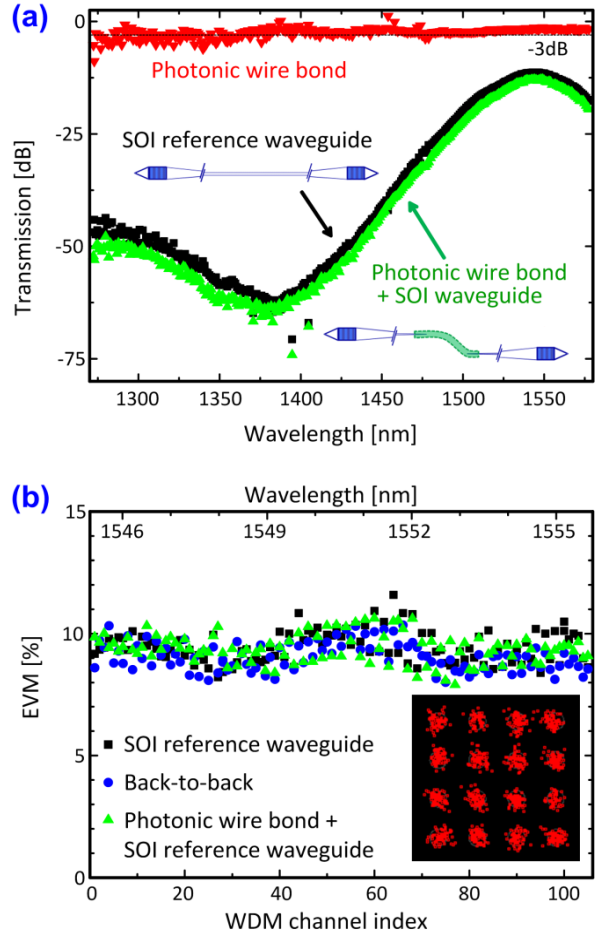

**Figure 3:** Optical characterization and experimental testing of a photonic wire bond (PWB) prototype **(a)** Transmission characteristics: The lower curve ('Photonic Wire Bond + SOI waveguide', green) represents the transmission through the entire PWB assembly consisting of two grating couplers, two SOI waveguide sections and the PWB interconnect. To obtain the net insertion loss of the PWB (red), the measured transmission spectrum was normalized to that of an SOI reference structure comprising only two grating couplers and an SOI waveguide (black). **(b)** Data transmission using wavelength division multiplexing (WDM): Each of the 106 carriers was modulated with a 16 QAM signal at a symbol rate of 12.5 GBd. The error-vector magnitude (EVM) was measured for the back-to-back case, the transmission through the SOI reference waveguide, and the transmission through the PWB assembly. Inset: 16 QAM constellation diagram of a single channel.

The predicted low-loss broadband transmission capabilities of photonic wire bonds were experimentally confirmed. For a test structure linking two 500 nm-wide SOI waveguides on the same chip, we measured the transmission in the entire wavelength range between 1240 nm and 1580 nm, see Figure 3 (a). The net insertion loss of the PWB was obtained by normalizing the transmission spectrum of the entire assembly to that of a reference SOI waveguide. The variations in the measured loss spectrum below 1480 nm are not caused by the PWB interconnect itself but by the measurement technique: The SOI grating couplers were optimized for an operation wavelength of 1550 nm with a 1 dB bandwidth of approximately 38 nm. For wavelengths outside that range, only residual scattered light was coupled between the fibers and the SOI waveguides, and the measurement data is subject to large uncertainties. Within the C-band of optical fiber communications (1530 nm - 1565 nm), the PWB features an essentially wavelength-independent total insertion loss of $(1.6 \pm 0.13)$ dB, which corresponds to $(0.8 \pm 0.1)$ dB per chip-to-PWB interface. For the entire wavelength range between 1240 nm and 1580 nm, the total insertion loss amounts to $(2.5 \pm 1.1)$ dB, which is equivalent $(1.3 \pm 0.8)$ dB per interface. Extrapolating the transmission spectrum to wavelengths beyond 1580 nm, we estimate the 1 dB bandwidth to be larger than 300 nm. PWB clearly outperforms high-efficiency fiber-chip grating couplers [6] and can well compete with planar coupling schemes fabricated by advanced electron beam lithography [14].

Photonic wire bonds can handle multi-Terabit/s data streams comprising tens of wavelength channels, even if phase-sensitive advanced modulation formats are used. We experimentally confirm the absence of transmission impairments through reflection, polarization mode dispersion, or nonlinearities. A transmitter setup similar to the one described in ref. [23] was used to generate an aggregate data stream of 5.25 Tbit/s. This stream consists of 105 carriers each of which is modulated with a 16 QAM signal at a symbol rate of 12.5 GBd, see Methods section. At the receiver side, the error-vector magnitude (EVM) of the individual wavelength channels is analyzed. Figure **3** (b) depicts the results for the back-to-back case (BTB), the transmission through the SOI reference waveguide, and the transmission through the PWB assembly. No EVM penalty could be measured for the transmission through the PWB compared to the reference waveguide and the BTB case. The low insertion loss of the PWB allows to keep the launched on-chip signal power low and hence prevents signal impairments from nonlinearities in the nanophotonic SOI waveguides. To the best of our knowledge, the demonstrated 5.25 Tbit/s is the highest data rate that was ever transmitted through an SOI nanowire. The data rate was only limited by the available transceiver equipment, not by the PWB or the SOI waveguide. Assuming a 5 μm pitch between neighbouring PWB waveguides, our concept enables unprecedented interconnect densities of at least 1 Pbit/s/mm along the chip edge (200 Pbit/s/mm$^2$ on the chip surface). This is more than two orders of magnitude larger than recently demonstrated values [16].

In summary, we have introduced photonic wire bonding as a novel concept for single-mode chip-to-chip interconnects. Direct-write two-photon lithography is deployed to fabricate 3D freeform waveguides. For the first time, we demonstrate single-mode PWB bridges between nanophotonic silicon-on-insulator waveguides. Our structures realize highly efficient chip-to-chip coupling with insertion losses of (2.5 ± 1.1) dB between 1240 nm and 1580 nm, and with losses of (1.6 ± 0.1) dB within the infrared telecommunication C-band (1530 – 1565 nm). In a proof-of-principle experiment, a wavelength-division multiplexing (WDM) signal with 5.25 Tbit/s aggregate data rate was transmitted through a photonic wire bond without any measurable signal degradation. Photonic wire bonding paves the path towards interconnects with spatial densities in the Pbit/s/mm range. Enabling efficient co-integration of different photonic device technologies, we believe the concept to be of groundbreaking nature for large-scale photonic integration.

## Methods

### Modeling

Simulations of the transition between the integrated SOI waveguide and the PWB section were performed using a commercially available fully vectorial time-domain solver (CST Microwave Studio). The modeled taper structure is depicted in Figure 1 (b). The SOI waveguide consists of a 500 nm wide and 220 nm high silicon strip (refractive index $n_{Si}$ = 3.48 at 1550 nm) deposited on a silicon dioxide buffer layer ($n_{SiO2}$ = 1.44 at 1550 nm). The PWB waveguide core consists of a polymer material (SU-8, $n_{SU8}$ = 1.57 at 1550 nm) and features a rectangular cross section of 1.4 μm width and 1 μm height. In the transition region, the SOI waveguide width is tapered down to a tip width $w_{tip}$ between 10 nm and 100 nm over a length of 30 μm. The inverse SOI taper is surrounded by a tapered section of the PWB polymer waveguide with an initial height of 450 nm, an initial width of 760 nm and a length of 20 μm. The PWB waveguide core is covered by a low-index cladding (e.g., Cytop, $n_{Cy}$ = 1.34 at 1550 nm), which is not depicted in Figure 1 (b) for the sake of clarity. The geometrical parameters are obtained from a few systematic trials, but they do not represent the optimum taper geometry. Further improvement of the coupling efficiency is possible by systematic numerical optimization.

### Sample fabrication

SOI waveguide structures were fabricated on a 200-mm CMOS line using 193 nm deep-ultraviolet (DUV) lithography and chlorine-based reactive ion etching [24]. The thickness of the SOI device layer (waveguide height) was 220 nm, and the buried oxide was 2 μm thick.

We fabricated our PWB prototypes with a commercially available 3D laser lithography system based on an inverted microscope (Photonic Professional, Nanoscribe GmbH). The system uses a pulsed laser beam at 780 nm wavelength with approximately 100 MHz repetition frequency and 150 fs pulse width. The lateral positions of the photonic chips were visually determined from the wide-field image taken by the microscope camera through the lithography objective. The vertical position was adjusted manually by optimizing the focus of the image. Achievable alignment tolerances are estimated to be better than 500 nm in the horizontal and better than 1 μm in the vertical direction. We expect that more precise alignment using machine vision techniques will reduce the PWB insertion losses further. MicroChem SU-8 2075 photoresist (ref. index $n$ = 1.57 at 1550 nm) was used for the PWB waveguide core. After development of the exposed resist, the structure was immersed in index matching liquid (Cargille Laser Liquid, code 3421, $n$ = 1.30 at 1550 nm) to emulate a low-index cladding material. Some remaining index matching liquid is visible in Figure 2 (b).

The PWB depicted in Figure 2 (b) bridges a distance of approximately 100 μm between the two SOI waveguide tips and compensates a lateral displacement of the photonic chips of approximately 25 μm in the horizontal and approximately 12 μm in the vertical direction. Fourth-order polynomials are used to define the 3D PWB trajectory connecting the two SOI waveguide tips while ensuring kink-free transitions between the SOI waveguide and the PWB section. The PWB apex reaches a height of 18 μm above the top surface of the upper chip (Chip 2). The samples were stored at normal atmosphere, and no special measures are taken to protect the waveguide against environmental influences such as oxygen or humidity. Repeated testing of the waveguides over several weeks with optical powers of up to 100 mW did not lead to any degradation of the transmission properties. The mechanical stability of the structures is excellent – free-standing waveguide arcs with less than 2 μm diameter can easily span distances of more than 100 μm and are not affected by manual handling of the chip with tweezers or by intensive rinsing in water after the development step. Moreover, the wire bonds exhibit strong adhesion to the silicon chip surface and do not detach even when treated with oxygen plasma or when immersed in acetone.

**Optical characterization**

To obtain the wavelength-dependent insertion loss of the PWB interconnects, we measured the transmission through the entire assembly consisting of two grating couplers, two SOI waveguide sections and the PWB itself, see Figure 2 (a). The transmission spectrum was then normalized to that of a reference structure comprising only two grating couplers and an SOI waveguide. Naturally, this procedure is only valid if the SOI waveguide sections of the PWB assembly are identical to the SOI reference waveguide. For the chip-to-chip PWB interconnect depicted in Figure 2 (b), the silicon waveguides are tapered down from 500 nm to 100 nm over a length of 100 μm. These structures were originally designed for direct coupling of single-mode fibers to SOI waveguides. In our experiment, however, only the last 20 μm of the inverse taper were embedded into the PWB polymer waveguide, such that additional losses occurred in the unused taper sections of the SOI feed waveguides. As a reference, only a 500 nm wide SOI waveguide without taper and hence with lower propagation loss was available. Using this structure, we obtain a strongly wavelength-dependent and significantly overestimated insertion loss of the PWB [17]. In contrast to that, the transmission spectrum depicted in Figure 3 (a) was obtained from a PWB interconnect between two 500 nm wide SOI waveguides on a common substrate. For these structures, a reference waveguide of the same width was available on the very same chip, which reduced the errors introduced by different waveguide widths and fabrication processes. Results similar to the one depicted in Figure 3 (b) were reproduced with different PWB samples.

**Transmission experiments**

For the transmission experiment, a setup similar to the one in ref. [23] was used to generate a wavelength-division multiplexing (WDM) signal which comprises 105 wavelength channels between 1543.42 nm and 1555.74 nm. The optical carriers were derived from a single mode locked laser (MLL, Ergo-XG, Time-Bandwidth Products) operating at a repetition frequency of 12.5 GHz. The output spectrum of the MLL was broadened in a highly nonlinear fiber, and a first wavelength-selective switch (WaveShaper, Finisar) was used to shape the spectrum [23]. Carriers belonging to even and odd channels are separated in an optical disinterleaver and subsequently encoded with a 16-state quadrature-amplitude modulation (QAM) format. Bandwidth-limited sinc-shaped Nyquist pulses are used to enable dense spectral packing of the WDM channels . An erbium-doped fiber amplifier (EDFA) compensates for the coupling losses to the integrated SOI waveguides. For the transmission through the PWB assembly and through the SOI reference waveguide, 13 dBm of optical power was used in the input fiber before the device under test. At the receiver side, individual WDM channels are isolated in a second wavelength-selective switch. The signal is then demodulated and characterized with an optical modulation analyzer (Agilent N4391A) with a tunable laser serving as local oscillator. An optical attenuator is used to keep the input power into the second wavelength-selective switch at a constant value of 0 dBm. A typical constellation diagram is depicted in the inset of Figure 3 (b); it refers to the transmission through the PWB and was measured at the carrier closest to 1550 nm. As a quantitative measure of the signal quality, we use the error vector magnitude (EVM), which describes the effective distance of a received complex symbol from its ideal position in the constellation diagram. The EVM can be interpreted as an extension of the Q-factor technique to the case of higher-order modulation formats and is directly connected to the bit-error ratio (BER) if the signal is impaired by additive white Gaussian noise [25]. For the back-to-back case, we found an average EVM of 9.1 %, which corresponds to a BER of approximately $0.9 \times 10^{-4}$. For the transmission through the SOI reference waveguide the average EVM is only slightly increased to 9.5 % (BER = $1.6 \times 10^{-4}$), and the transmission through the PWB yields essentially the same value (EVM = 9.4 %, BER = $1.4 \times 10^{-4}$). The PWB does hence not introduce any EVM penalty and the BER for all carriers is below the threshold of $2.3 \times 10^{-3}$ ($1.9 \times 10^{-2}$) for state-of-the-art (advanced) forward-error correction (FEC) [25], [26].

# Acknowledgements

This work was supported by the Center for Functional Nanostructures (CFN) of the Deutsche Forschungsgemeinschaft (DFG) (project A 4.7), by the Initiative of Excellence at Karlsruhe Institute of Technology (KIT), by the Karlsruhe Nano-Micro Facility (KNMF), by the Karlsruhe School of Optics & Photonics (KSOP), by the European Research Council (ERC Starting Grant 'EnTeraPIC', number 280145), and by the German Federal Ministry of Economics and Technology (BMWi) within the project POLINA. Silicon-on-insulator waveguides were fabricated by the European silicon photonics platform ePIXfab (www.epixfab.eu).

# Notes

The authors declare that they have no competing financial interests. Correspondence and requests for material should be addressed to C. K. (e-mail: christian.koos@kit.edu)